# Calculation of geometrical and spin features of a series of metal-endofullerenes

V.S. Gurin

*Physico-Chemical Research Institute,*

*Belarusian State University, Leningradskaya str., 14, Minsk, 220080, Belarus;*

*E-mail: gurin@bsu.by; gurinvs@lycos.com*

A series of endofullerenes $M@C_{60}$ were calculated from the first principles (unrestricted Hartree Fock and DFT B3LYP methods) with effective core potential (M = Ag, Cs) and all-electronic basis set (M = Li, Na, Cu). An arbitrary symmetry distortion (down to $C_1$ point group) was assumed. The geometrical and electronic properties of $M@C_{60}$ are compared for this series of endoatoms including the off-center position of endoatoms within $C_{60}$, effective charges and spin density. The latter values are featured for $Ag@C_{60}$ and $Cu@C_{60}$.

*Keywords*: endofullerenes, ab initio calculations, metal atoms, quantum computing

## 1. Introduction

Endohedral fullerenes M@$C_{60}$ are of great interest from the point of both experimental and theoretical studies of new type of chemical species [1-3]. Their formation occurs due to the geometrical confinement of atoms inside the closed carbon cage $C_{60}$. A question on chemical bonding of an endoatom with carbon *a priori* is open. Since the simple geometrical reasons provide the possibility of these systems. However, bare metal atoms as well non-metal elements in atomic form usually are strongly reactive, and the M-C bond formation is very probable. It is known that non-metal atoms such as nitrogen and phosphorus can produce the endohedral structures M@$C_{60}$ with negligible effect upon the carbon cage [4,5], while the active metals (alkali, rare earth elements) interact much stronger with $C_{60}$ [2,3,5,6]. Weakly interacting atoms are localized in the center of $C_{60}$ molecule, and the electronic state of these atoms deserves a special study [7]. The state of metal atoms inside fullerenes as $C_{60}$ as well in the higher ones, $C_n$, n>60, has been calculated in many works (see review in Refs. 2,3). Different heavy elements have been considered as endoatoms within $C_{60}$, e.g. K, Fe, Mn, Gd [8-11], and principal possibility of M@$C_{60}$ exists for various M. From geometrical reasons all monoatomic endofullerenes are admissible, however, experimental evidences exist for limited number of atoms, mostly, active metals which interact strongly with the carbon cage, and non-metal elements (N, P, H) those appear to be weakly interacting. For theoretical evaluation the symmetrical structures were considered in more degree as far as the original $C_{60}$ molecule possesses the icosahedral symmetry. In the present work, we calculate a series of M@$C_{60}$ models assuming arbitrary symmetry distortion to demonstrate how the nature of endoatoms influences the properties of the endofullerenes. M@$C_{60}$ with M=Li, Na, Cs are really existing species [12-14] while the species with M=Ag and Cu are not well established yet [15,16], however, there are arguments in support of their existence [17]. This set of metals considered as endoatoms belongs to the 1[st] group of the Periodic System of Elements covering the examples from both Ia and Ib parts. This study can shed a light on the nature of endofullerenes stability and their geometrical features.

## 2. Calculation method

For calculations we used the *ab initio* SCF unrestricted Hartree-Fock (HF) and DFT calculations. Within the DFT method electron correlation effects were included by selection of commonly used B3LYP functional that is considered as good balanced one for different molecules. Basis sets were combined from all-electronic and effective core potential (ECP) versions: for Ag the 28-electron core and for Cs the 46-electron one were used [18], for Cu – the basis of the 6-31G* quality, for C - STO-3G and for Li and Na both STO-3G and 6-31G were taken for calculations.

This selection of the rather simple basis sets is taken because considerable calculation task with 60 carbon and other heavy atoms and allowance of asymmetry of the models. Calculation results at the present levels should be considered as simulation of general trends in features of these molecules rather than exact quantitative evaluation of numerical data to compare with experiment. At present, besides calculations of endofullerenes at the semiempirical levels, HF and DFT [2,3,7-11], the higher-level approaches were performed to account the electron correlation effects with more accuracy, however, the problem is far from completion. The local orbital approach has been used for Na-$C_{60}$ system with the complete active space configuration interaction (CAS-CI) method, and the local approach made it possible to reduce computation tasks [19,20]. However, the highly delocalized π-electronic system is of principal importance for the bonding of $C_{60}$ cage with metals (both in endo- and exo-structures), and this simplification cannot be considered as completely satisfying.

The doublet electronic states were considered here for all the models of M@$C_{60}$ since the ground state of $C_{60}$ is known to be closed shell singlet. M@$C_{60}$ structures were built by adding one metal atom (with open electronic shell) inside $C_{60}$. The geometry optimization of the structures was done at the two levels: (i) with conservation of initial symmetry constraints ($I_h$) by placing endoatoms into the center and (ii) under arbitrary distortion down to $C_1$ point group that allowed any off-center shift of endoatoms. A distortion of the carbon cage was also quite feasible and really occurred in small degree. We indicate this distortion through the minimum and maximum of C-C bond lengths (Table 1).

NWChem software (versions 4.3-4.5) [21] was utilized for the calculations, and the above basis sets were taken from this package without modifications.

## 3. Results and Discussion

The calculation results for geometry of the models, binding energies, effective charges at endoatoms and spin densities derived from the Mulliken occupancy analysis are summarized in Table 1. Fig. 1 displays the geometry of the structures taking into account the off-center shifts obtained by the calculations. This shift occurs in all cases, but rather different for the metal atoms under study. The endoatoms in minimum-energy optimized structures are located not along the symmetry axes of the original $C_{60}$ molecule ($C_2$, $C_3$, and $C_5$). More detailed analysis requires complete study of potential energy surfaces for endoatoms with each metal and not given here. We note only the explicit asymmetry of the models that follows both from HF and DFT calculations. This fact is not surprising knowing numerous theoretical research on metal-endofullerenes [1-3,5,6]. Its origin may be associated with familiar symmetry distortion phenomena in metal complexes, like the Jahn-Teller effect [22]. This indicates the strong interaction of metal atoms with the carbon cage. Our calculation results demonstrate rather significant changes both in geometry and electronic structure of the M@$C_{60}$ models as compared with original $C_{60}$ counterparts.

The data presented in Table 1 show that the off-center shift depends strongly on the nature of metal atom inside the fullerene cage. The noticeable more shift occurs for alkali metal atoms (besides Cs) than for silver and copper. In the case of Cs, the less effect can be due to the larger atomic (and ionic) radius that geometrically prevents the possible relocation of Cs from the central position.

Effective charges at the endoatoms are rather featured not only in their values but also in the signs. The big positive charges for Na and Cs are easily understandable as familiar property of the metal atoms to be an electronic donor. For Li there appears the difference in sign of effective charges in HF and DFT results. Possibly, the strong mobility of Li atom due to its small radius provides the variance in the calculation results at the different level, and the theory for Li-endostructures requires more deep analysis and upgrade. Within the framework of present data, the positive charge by DFT seems to be more reliable and fits in the consistence with another recent calculations at the DFT level [23,24]. Thus, the accounting of electron correlation occurring with the DFT calculations (B3LYP) is rather important for correct value of effective charges. The geometry is less sensitive with respect to this factor. The anomalous charge is observed also for silver, and both HF and DFT data give close results (Table 1). Ag atom is not efficient electronic donor and has the larger atomic radius than Li and Na. These results on effective charges can be understood by comparison of ionization potentials for a series of the metals under consideration. They vary in the sequence Ag > Cu > Li > Na > Cs. This sequence correlates with effective charges (besides, Li, see above), the effective charges vary from negative one to the superionization in the case of most electronic-donor atom, Cs.

Another interesting and novel results of these calculations are collected in the last column of Table 1 as the values of spin density in the optimized structures. Alkali metals indicate zero spin density at the endo-atom, while silver and copper demonstrate the values close to unity. That was supported by both HF and DFT data. The reason of such behavior of Ag and Cu can be associated with their electronic structure, in particular, with explicit contribution of *d*-orbital. In the case of alkali metals, including the heavy Cs, the main contribution occurs from *ns*-orbitals and for Cs *5p* ones become to be of importance. Alkali elements behave as the typical *s*-metals with no featured spin density. Thus, the spin properties of Ag@$C_{60}$ and Cu@$C_{60}$ can suggest a special interest for them, e.g. for construction of quantum computing elements. The endofullerenes with non-metal elements in which the spin of endoatoms is stabilized by the carbon cage have been considered in such function [25-27], however non-metal endoatoms are weakly interacting with $C_{60}$. $Ag@C_{60}$ and $Cu@C_{60}$ are strongly bound species, and their spin state can be controlled by external biases through the cage. These aspects and quantum-computing models of endofullerenes are the subject of our future studies.

The binding energies for all models correspond to formation of possible stable structures, but the HF result, likely, overestimate the energies. Meanwhile, the rather large binding energy for $Ag@C_{60}$ is of interest, though, it cannot be compared with experiment yet. The comparison of calculation results derived by the HF approach and DFT method indicates that there are some differences, but the main trends in the features of models are similar (besides the above noted effective charge of Li atom). The DFT method demonstrates the less distortion of the carbon cage due to the presence of endoatoms. This is known feature of DFT calculations, and the data of HF approach give the more distorted structure about 0.05 Å. The distortion of the carbon cage is quite expectable for endostructures, and the calculations evidence that there is almost no difference for the metals under consideration. Within the DFT results this effect is less pronounced that is known as one of DFT features as compared with HF calculations. Experimental data, if available, can respond which series of results are of more reliability.

Direct experimental evidences for the geometrical characteristics of the metal-endofullerenes are not available at present, however, another calculations for $Na@C_{60}$ result in the difference of maximum and minimum bond lengths about 0.1 Å [8]. For $Li@C_{60}$ the large values of off-center shift of lithium atom was reported in Refs. 23,24, but values of the effective charges are rather variable depending on calculation level in these papers. Evidently, an accuracy of all calculations should be upgraded for these systems, however, the principal properties of endofullerenes are reproduced at the levels available now.

## 4. Conclusions

A series of endofullerenes $M@C_{60}$ with M=Li, Na, Cs, Cu, Ag were calculated at the HF and DFT levels assuming arbitrary symmetry distortion. All they are featured by the off-centre shift of endoatoms that is strongly dependent on M: the maximum (more than 1 Å) for Li and Na and minimum (about zero) for Cs. $Ag@C_{60}$ demonstrates the anomalous charge transfer (silver is negative), but the structure is stable. $Ag@C_{60}$ and $Cu@C_{60}$ are also characterized by the high values of spin densities for the endoatoms that has been suggested for models of quantum computing elements.

**Acknowledgements**

The authors want to acknowledge the support of this work due to participation in the project under the Ministry of Education of Belarus.

Table 1. Characteristics of the endofullerenes with a series of metal atoms calculated by the HF and DFT method

| Model | Minimum C-C distance, Å | Maximum C-C distance, Å | Off-centre position of metal atoms, Å | Binding energy of M-$C_{60}$ system, eV | Net charge of M atoms | Spin density at M atoms |
|---|---|---|---|---|---|---|
| HF results | | | | | | |
| $C_{60}$ | 1.38 | 1.46 | | | | |
| Ag@$C_{60}$ | 1.40 | 1.54 | 0.28 | 9.94 | -0.10 | 0.99 |
| Cu@$C_{60}$ | 1.40 | 1.54 | 0.48 | 2.8 | +0.33 | 0.98 |
| Li@$C_{60}$ | 1.41 | 1.54 | 1.10 | 8.35 | -0.15 | 0 |
| Na@$C_{60}$ | 1.41 | 1.54 | 1.30 | 13.19 | +0.85 | 0 |
| Cs@$C_{60}$ | 1.41 | 1.54 | 0.07 | 5.87 | +1.19 | 0 |
| DFT results | | | | | | |
| $C_{60}$ | 1.41 | 1.48 | | | | |
| Ag@$C_{60}$ | 1.41 | 1.48 | 0.20 | 1.67 | -0.16 | 0.97 |
| Cu@$C_{60}$ | 1.41 | 1.48 | 0.07 | 0.55 | +0.52 | 0.88 |
| Li@$C_{60}$ | 1.41 | 1.49 | 1.08 | 0.67 | +0.59 | 0 |
| Na@$C_{60}$ | 1.42 | 1.48 | 1.15 | 5.98 | +0.85 | 0 |
| Cs@$C_{60}$ | 1.41 | 1.48 | ~0 | 0.23 | +1.21 | 0 |

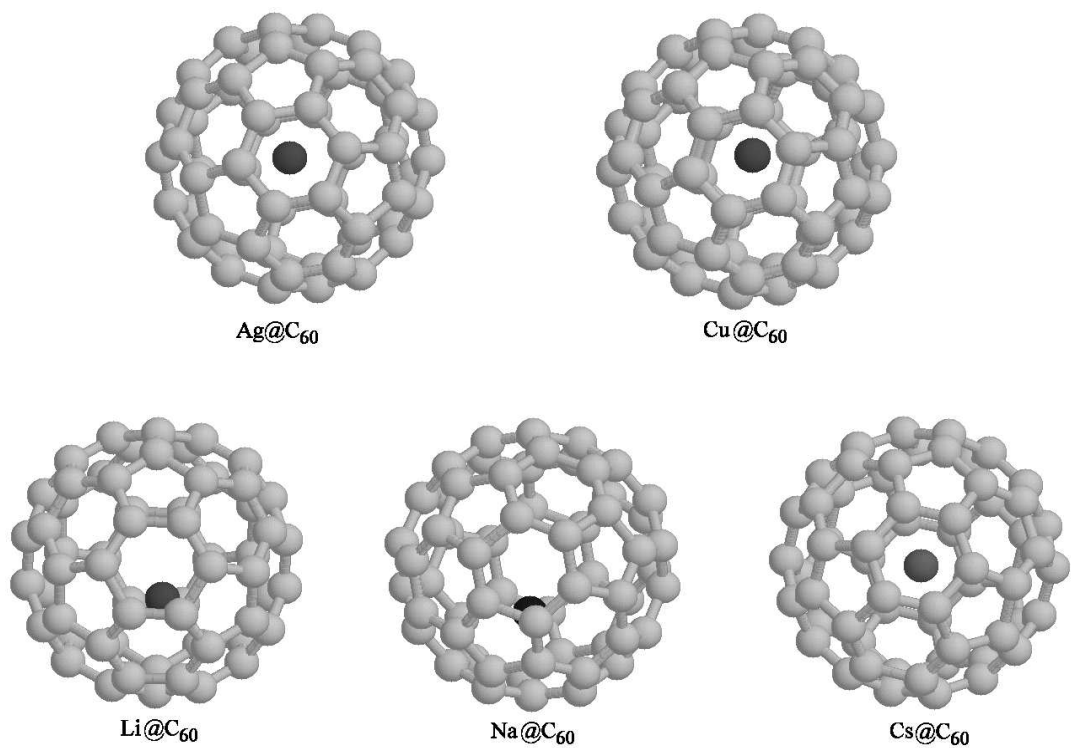

**Fig. 1.** The structures of M@C$_{60}$ endofullerenes with different metal atoms with the shift of endoatoms calculated.